\documentclass[twocolumn,showpacs,preprintnumbers,amsmath,amssymb,prl,nofootinbib]{revtex4}

\usepackage{graphicx}% Include figure files
\usepackage{dcolumn}% Align table columns on decimal point
\usepackage{bm}% bold math
\usepackage{times}
\usepackage{multirow}
\usepackage{footmisc}
\usepackage{color}

\begin{document}

\title{Tension and stiffness of the hard sphere crystal-fluid interface}

\author{A. H\"artel$^1$}
\author{M. Oettel$^2$}
\author{R. E. Rozas$^1$}
\author{S. U. Egelhaaf$^3$}
\author{J. Horbach$^1$}
\author{H. L\"owen$^1$}
% eMail soll am Besten mit Sternchen angezeigt werden. 
%\email{email-address}
\affiliation{%
$^1$ Institut f\"ur Theoretische Physik II: Weiche Materie, 
Heinrich-Heine-Universit\"at D\"{u}sseldorf, 
Universit{\"a}tsstra{\ss}e 1, D-40225 D\"{u}sseldorf, Germany\\
$^2$ Johannes-Gutenberg-Universit\"at Mainz, 
Institut f\"ur Physik, WA 331, D-55099 Mainz, Germany\\
$^3$ Condensed Matter Physics Laboratory,
Heinrich-Heine-Universit\"at D\"{u}sseldorf,
Universit{\"a}tsstra{\ss}e 1, D-40225 D\"{u}sseldorf, Germany}
\date{\today}

\begin{abstract}
A combination of fundamental measure density functional theory
and Monte Carlo computer simulation is used to determine the
orientation-resolved interfacial tension and stiffness for the equilibrium
hard-sphere crystal-fluid interface.  Microscopic density functional
theory is in quantitative agreement with simulations and predicts a
tension of $0.66\,k_{\rm B}T/\sigma^{2}$ with a small anisotropy of about
$0.025\,k_{\rm B}T$ and stiffnesses with e.g.~$0.53\,k_{\rm B}T/\sigma^{2}$ for
the (001) orientation and $1.03\,k_{\rm B}T/\sigma^{2}$ for the (111)
orientation. Here $k_{\rm B}T$ is denoting the thermal energy and $\sigma$
the hard sphere diameter. We compare our results with existing experimental findings. 
\end{abstract}


\maketitle

%68.08.De 	Liquid-solid interface structure: measurements and simulations 
%64.70.D- 	Solid-liquid transitions 
%05.20.Jj 	Statistical mechanics of classical fluids 
%82.70.Dd       Colloids
%61.50.Ah 	Theory of crystal structure, crystal symmetry; calculations and modeling, !!!!!!!!! besser herauslassen????

Solidification and melting processes involve crystal-fluid interfaces
that separate the disordered from the ordered phase. Understanding
the properties of such interfaces on a microscopic scale is pivotal to
control and optimize crystal nucleation and the emerging microstructure
of the material. Important applications include the fabrication
of defect-free metallic alloys \cite{antonowicz_rams18_2008} and of
photonic \cite{dziomkina_sm1_2005}, phononic \cite{gorishnyy_prl94_2005}
and protein \cite{sear_jpcm19_2007} crystals.  In equilibrium, i.e.\
between a coexisting crystal and fluid phase, creating a crystal-fluid
interface results in a free energy penalty per area that is called
interfacial tension.  Unlike the liquid-gas or fluid-fluid interface,
the structure  of the solid-fluid interface depends on its orientation
\cite{woodruff_book_1980}.  This anisotropy is associated with a
difference between the interfacial tension and the interfacial stiffness
of a crystalline surface.

Predicting crystal-fluid interfacial tensions by a molecular theory
is a very challenging task.  Classical density functional theory of
freezing provides a unifying framework to describe the solid and liquid
on the same footing and is therefore in principle a promising tool.
In this respect, the simple athermal hard sphere system which exhibits a
freezing transition from a fluid into a face-centered-cubic (fcc) crystal,
is an important reference system. The accuracy of previous density
functional calculations of the hard sphere solid-fluid interface
\cite{curtin_prl59_1987,curtin_prb39_1989,marr_pre47_1993,ohnesorge_pre50_1994},
however, was hampered by the lack of knowledge of a reliable functional
and severe restrictions in the parametrization of the trial density
profile.

\begin{figure}
%\center
\includegraphics[width=8.5cm]{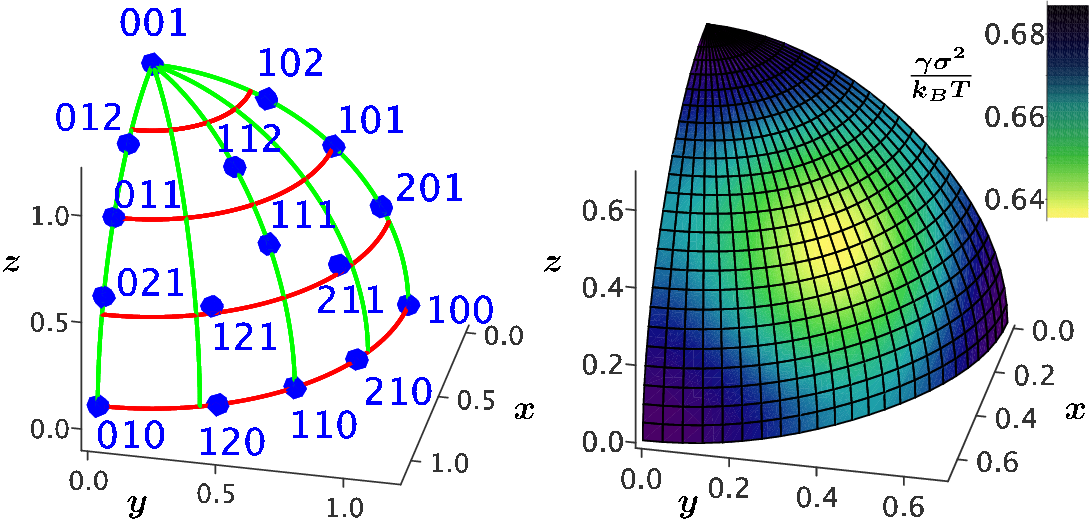}
\caption{\label{fig_orientations} 
(Color online) In the left panel, the surface orientations, as
listed in Table \ref{tab_values}, are indicated on an octant of the
unit sphere. The right panel shows a Wulff plot of the corresponding
interfacial tension $\gamma(\hat{n})$; here, the colors display the
value of the tension for a given orientation.}
\end{figure}

In this letter, interfacial tensions and stiffnesses of the equilibrium
hard sphere crystal-fluid interface are predicted using fundamental
measure density functional theory \cite{hansen-goos_jpcm18_2006}
which has been shown to predict accurate bulk freezing data
\cite{oettel_pre82_2010}. The interfacial tension and stiffness for
five different orientations are obtained, namely along the (001), (011),
(111), (012) and (112) orientations (see Fig.~\ref{fig_orientations}). A
small orientational anisotropy for the tensions is found and the average
tension is about  $0.66\,k_{\rm B}T/\sigma^{2}$ with $k_{\rm B}T$
denoting the thermal energy and $\sigma$ the hard sphere diameter. For
the stiffnesses the data are spread in a much wider range between
$0.28\,k_{\rm B}T/\sigma^{2}$ for the (011) orientation with lateral
direction [$\bar{1}00$] and $1.03\,k_{\rm B}T/\sigma^{2}$ for the
(111) orientation. We have also conducted Monte Carlo simulations to
extract the stiffness from capillary wave fluctuations for the above
orientations except (012), thereby improving the accuracy of earlier data
\cite{davidchack_jcp108_1998,davidchack_prl85_2000,mu_jpcb109_2005,davidchack_jcp125_2006,amini_prl97_2006,zykova-timan_jpcm21_2009,davidchack_jcp133_2010}.
We find quantitative agreement between density functional theory and
computer simulation.

In equilibrium, the athermal hard sphere model system is solely
characterized by the volume fraction $\phi$; the thermal energy
$k_{\rm B}T$ just sets the energy scale. The fluid-solid(fcc) freezing
transition is first-order with coexisting fluid and solid volume
fractions of $\phi_{\rm f}=0.492$ and $\phi_{\rm s}=0.545$, respectively,
and a coexistence pressure of $p_c=11.576\,k_{\rm B}T/\sigma^3$
\cite{zykova-timan_jcp133_2010}. For a given volume $V$ containing
coexisting bulk fluid and solid, and a planar fluid-solid interface
of area $A$, the excess grand free energy per area is the surface or
interface tension, given by $\gamma=(\Omega+pV)/A$, with $\Omega$
denoting the grand-canonical free energy. For crystal-fluid interfaces,
$\gamma$ depends on the orientation of the interface, characterized by a
normal unit vector $\hat{n}$ relative to the crystal lattice. The latter
is fixed with the fcc cubic unit cell edges parallel to the Cartesian
coordinate axes of our system, see Fig.~\ref{fig_orientations}. 

The central quantity to describe thermal fluctuations, i.e.~capillary
waves, along the anisotropic crystal-fluid interface is the interfacial
stiffness defined tensorially \cite{fisher_prl50_1983} as
\begin{equation}
\bar{\gamma}_{\alpha\beta}(\hat{n}) 
=\gamma(\hat{n})
+  \frac{\partial^2\gamma(\hat{n})}{\partial\hat{n}_{\alpha}\partial\hat{n}_{\beta}}
\label{eqn_stiffness}
\end{equation}
for two directions $\hat{n}_{\alpha}$ and $\hat{n}_{\beta}$ that are
orthogonal to $\hat{n}$.

We calculate the tension of the hard sphere crystal-fluid interface using
classical density functional theory (DFT) that provides direct access
to the grand-canonical free energy $\Omega$ \cite{roth_jpcm22_2010}. We
employ the geometric fundamental measure approach first established
by Rosenfeld \cite{rosenfeld_prl63_1989,tarazona_prl84_2000} and
most accurately refined in the so-called White Bear version mark II
\cite{hansen-goos_jpcm18_2006}. A free minimization of this theory in
the bulk phases \cite{oettel_pre82_2010} gives accurate hard sphere
bulk coexistence data which are needed as a reliable input for the
calculation of interfacial tensions.  The
crystal-fluid phase transition occurs at a coexistence chemical potential
$\mu_{\rm c}/k_{\rm B}T=16.3787$ and a coexistence pressure $p_{\rm
c}\sigma^{3}/k_{\rm B}T=11.8676$. The coexistence packing fractions of
the fluid and solid are respectively $\phi_{\rm f} = 0.495$ and $\phi_{\rm
s} = 0.544$, in close agreement with the aforementioned computer simulation 
data \cite{zykova-timan_jcp133_2010}.

At the prescribed coexistence chemical potential $\mu_{\rm c}$, the
grand free energy functional is numerically minimized inside a rectangular 
cuboid box of lengths $L_{x}$, $L_{y}$, and $L_{z}$ with periodic boundary
conditions in all three directions \cite{ohnesorge_pre50_1994}. The
surface normal is pointing along the $z$-direction and the box length
$L_{z}$ is chosen large enough (about $50-60\sigma$) to ensure a large
part of bulk crystal and fluid phase at coexistence which are separated
by two interfaces. The lateral dimensions $L_{x}$ and $L_{y}$ of the
box depend on the surface orientation relative to the fcc crystal. They
are determined by the minimal size of a periodic rectangular cross
section which accomodates the prescribed relative orientation.
The density field is resolved on a fine rectangular grid in real space
with a spacing of about $0.02\sigma$. Starting from an initial profile
which contains the two bulk parts of pre-minimized crystal and fluid,
the density functional is minimized using a Picard iteration scheme
combined with a {\it direct inversion in the iterative subspace}
method \cite{pulay_cpl73_1980,kovalenko_jcompc20_1999} and a simulated
annealing technique \cite{ohnesorge_pre50_1994}. Finite size effects due
to the finite grid size were excluded by also using smaller grid spacings
to ensure free minimization of the density functional in practice.

\begin{figure}
\includegraphics[width=8.0cm]{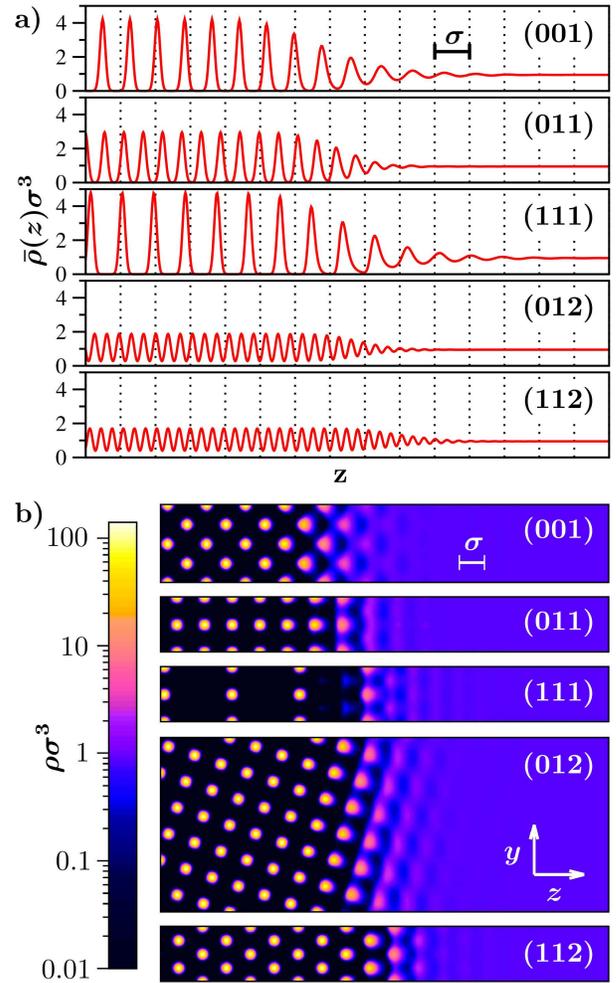}
\caption{\label{fig_density-profiles} 
(Color online) DFT results: a) Laterally integrated density profiles
${\bar \rho} (z)$ for the five surface orientations, as indicated; b)
contour plots at $x=0$. The periodic length of the total profiles in
$z$-direction is $50.15\sigma$ (001), $53.19\sigma$ (011), $65.15\sigma$
(111), $56.07\sigma$ (012), and $61.42\sigma$ (112).}
\end{figure}

Results for the minimized density profiles are displayed in
Fig.~\ref{fig_density-profiles} for five different orientations. Both
the laterally integrated ($z$-resolved) density field ${\bar \rho} (z) = \frac {1}
{L_{x}L_{y}} \int_0^{L_{x}}\int_0^{L_{y}}\rho(x,y,z) dy dx$ and contour
plots, $\rho(x=0,y,z)$, are shown.

The DFT results for the interfacial tension are summarized in Table
\ref{tab_values} for five different orientations. With a slight
orientational dependence, all the values vary around $0.66\,k_{\rm
B}T/\sigma^{2}$. The errors given in Table \ref{tab_values} are
estimated from several independent minimization runs. Since
the anisotropy is weak, the orientational resolved interfacial
tension can be well-fitted by the cubic harmonic expansion
\cite{fehlner_cjp54_1976,rozas_epl93_2011}
\begin{eqnarray}
\frac{\gamma(\hat{n})}{\gamma_0} &=& 
1+\epsilon_1\left(Q-\frac{3}{5}\right) 
+ \epsilon_2\left(3Q+66S-\frac{17}{7}\right) \notag \\
& & +\epsilon_3\left(5Q^2-16S-\frac{94}{13}Q+\frac{33}{13}\right)
\label{equ_gamma_expansion}
\end{eqnarray}
with $\hat{n}=(n_1,n_2,n_3)$, $Q=n_1^4+n_2^4+n_3^4$, $S=n_1^2n_2^2n_3^2$
and four fit parameters $\gamma_{0}$, $\epsilon_{1}$, $\epsilon_{2}$,
$\epsilon_{3}$. The expansion (\ref{equ_gamma_expansion}) can be
used to obtain the interfacial stiffness (\ref{eqn_stiffness})
from the DFT data of the anisotropic interfacial tension
\cite{zykova-timan_jcp133_2010,rozas_epl93_2011}. The
resulting anisotropy of the stiffness is considerably larger
than the one of the tension (thus, the resulting errors of
$\bar{\gamma}_{\alpha\beta}(\hat{n})$ are also larger).  For the
five different orientations considered in this work, the resulting data
for the interfacial stiffness and for the fit parameters are listed in
Table \ref{tab_values}.

In the Monte Carlo (MC) simulations, similar to the procedure in
\cite{zykova-timan_jpcm21_2009,zykova-timan_jcp133_2010}, inhomogeneous
hard-sphere systems at the coexistence pressure $p_{\rm c}$ are prepared
followed by production runs in the canonical ensemble. The canonical
MC simulation consists of particle displacement moves according to a standard
Metropolis criterion where the trial displacements of the particles
are randomly chosen from the interval [-0.11\,$\sigma$,+0.11$\sigma$].
The inhomogeneous solid-fluid systems are placed in rectangular cuboid simulation 
boxes of nominal size $L\times L \times 5L$ ($L\approx25\,\sigma$),
containing about $10^5$ particles. We apply periodic boundary conditions
in all three dimensions, the fcc crystal with $z$-extension of about
$2L$ resides in the middle of the box and is separated from the fluid
by two planar interfaces. Since the system is in equilibrium, the amount 
of crystal and fluid phase as well as the interfaces remain stable during 
the simulation. We consider the crystal orientations (001),
%The inhomogeneous solid-fluid systems are placed in cuboid simulation
%boxes of nominal size $L\times L \times 5L$ ($L\approx25\,\sigma$),
%containing about $10^5$ particles. We apply periodic boundary conditions
%in all three dimensions, the fcc crystal with $z$-extension of about
%$2L$ resides in the middle of the box and is separated from the fluid
%by two planar interfaces. We consider the crystal orientations (001),
(011), (111) and (112), see Fig.\ \ref{fig_orientations}. At each
orientation, 10 independent runs are performed and in each of these runs,
10,000 independent configurations are generated that are used for the
analysis of the interfaces.

The stiffnesses $\bar{\gamma}$ are extracted from the capillary wave
spectrum $\langle h^2(\vec{q}) \rangle$ \cite{rozas_epl93_2011},
i.e.~the correlation function of the interface height fluctuation
$h(\vec{q})$ (with $\vec{q}=(q_x,q_y)$ the two-dimensional wave-vector
along the lateral extension of the interface). In order to determine
$h(\vec{q})$ a criterion has to be introduced according to which one can
distinguish between fluid and crystal particles. Following work 
\cite{zykova-timan_jpcm21_2009,zykova-timan_jcp133_2010}, 
the rotational-invariant bond-order parameter $q_6q_6(i)$ was 
used \cite{steinhardt_prb28_1983,tenwolde_prl75_1995}. 
To distinguish between crystalline and fluid particles, 
we adopt the same criterion as in Refs. 
\cite{zykova-timan_jpcm21_2009,zykova-timan_jcp133_2010}, where a 
particle $i$ was identified as one with crystalline order if $q_6q_6(i)
> 0.68$, otherwise it was defined as a liquid-like particle. Moreover,
the local position of the interface is defined by the set of crystalline
particles at the interface (particles which have less than 11 crystalline
neighbors). Some particles in the liquid bulk identified
as crystalline were removed by searching the largest cluster among
the particles identified as interface-particles. The fluctuation of the local
interface position is defined as $h(x_i, y_i) = z_i - \langle z \rangle$,
with $i$ the index of a particle on the surface and $\langle z \rangle$
the instantaneous average location of the interface. The irregularly
distributed particle coordinates $(x_i, y_i)$ are then mapped onto a
regular grid in the $xy$ plane with grid spacing $\Delta x = \Delta y =
\sigma$ using Shepard interpolation \cite{rozas_epl93_2011}. Finally, the
height fluctuation $h(\vec{q})$ is obtained from a Fourier transformation
of the interpolated heights.

\begin{figure}
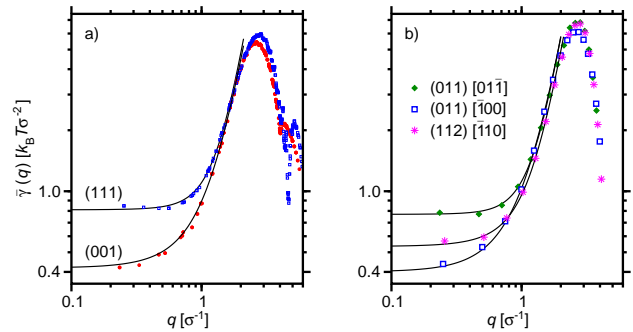

%\center
\includegraphics[width=4.0cm]{stiffness_mchs3_a.eps}
\hspace*{0.1cm}
\includegraphics[width=4.0cm]{stiffness_mchs3_b.eps}
\caption{\label{fig_stiffness} 
(Color online) $q$-dependent interfacial stiffness $\bar{\gamma}(q)$
for the (001) and (111) orientation [a)] as well as the (011) and (112)
orientation for the indicated directions [b)]. Note that for (112) only
the [$\bar{1}$10] direction is shown because $\bar{\gamma}(q)$ for
the [11$\bar{1}$] direction is very similar to that of the latter 
direction.
}
\end{figure}
Figure \ref{fig_stiffness} shows the $q$-dependent interfacial
stiffness, as defined by the equation $\bar{\gamma}_1(q_x)
q_x^2 + \bar{\gamma}_2(q_y) q_y^2 = k_{\rm B}T/[L_x L_y \langle
h^2(\vec{q})\rangle]$: for the (001) and (111) orientation
$\bar{\gamma}(q)=\bar{\gamma}_1(q_x)=\bar{\gamma}_2(q_y)$ holds
whereas for the (011) and (112) orientation there are two different coefficients
$\bar{\gamma}_1(q_x)$ and $\bar{\gamma}_2(q_y)$ that can be determined
from the latter equation by considering $q_y=0$ or $q_x=0$, respectively.
The solid lines in Fig.~\ref{fig_stiffness} are fits of the data for
$q<1.5\;\sigma^{-1}$ with the function $\bar{\gamma}(q)= \bar{\gamma}
+ a q^2 + b q^4$ yielding the values for the stiffness $\bar{\gamma}$
for $q\to 0$.  

\begin{table}
\caption{\label{tab_values}
Interfacial tensions $\gamma$ and stiffnesses $\bar{\gamma}$ in units 
of $k_{\rm B}T/\sigma^2$ for different surface normal vectors (round brackets) 
and tangential directions (square brackets). 
In DFT the tensions are measured directly, in the simulation the stiffnesses. 
The other data are listed italicized and are calculated using the fit 
function (\ref{equ_gamma_expansion}). The fit parameters are obtained 
from a least square fit to the measured data. For DFT they are 
$\gamma_0=0.664(2)k_{\rm B}T/\sigma^2$, $\epsilon_1=0.1076(120)$, 
$\epsilon_2=-0.01364(292)$, $\epsilon_3=-0.0023(209)$ and for 
simulation $\gamma_0=0.618(11)k_{\rm B}T/\sigma^2$, $\epsilon_1=0.0905(32)$,
$\epsilon_2=-0.00547(44)$, $\epsilon_3=0.0054(25)$. 
As a reference previous simulation results for 
tensions \cite{davidchack_jcp133_2010} and stiffnesses 
\cite{davidchack_jcp125_2006} are shown in the last column. 
The numbers in parentheses indicate the uncertainty in the last digit(s).}
\begin{ruledtabular}
\begin{tabular}{llccc}
 & orientation & theory & simulation & 
Ref.~\cite{davidchack_jcp125_2006,davidchack_jcp133_2010} \\
\hline
$\gamma$ & (001) & $0.687(1)$ & ${\it 0.639(11)}$ & $0.5820(19)$ \\
$\bar{\gamma}$ & (001) & ${\it 0.53(14)}$ & $0.419(5)$ & $0.44(3)$ \\
\hline
$\gamma$ & (011) & $0.665(1)$ & ${\it 0.616(11)}$ & $0.5590(20)$\\
$\bar{\gamma}$ & (011)$[\bar{1}00]$ & ${\it 0.283(35)}$ & $0.401(5)$ & $0.42(3)$\\
$\bar{\gamma}$ & (011)$[01\bar{1}]$ & ${\it 0.86(14)}$ & $0.769(5)$ & $0.70(3)$\\
\hline
$\gamma$ & (111) & $0.636(1)$ & ${\it 0.600(11)}$ & $0.5416(31)$ \\
$\bar{\gamma}$ & (111) & ${\it 1.025(79)}$ & $0.810(5)$ & $0.67(4)^a$ \\
\hline
$\gamma$ & (012) & $0.674(5)$ & ${\it 0.623(11)}$ & $0.5669(20)$ \\
$\bar{\gamma}$ & (012)$[\bar{1}00]$ & ${\it 0.454(57)}$ & $0.575(5)$ & $0.59(3)$ \\
$\bar{\gamma}$ & (012)$[02\bar{1}]$ & ${\it 0.71(12)}$ & $0.420(5)$ & $0.43(3)$ \\
\hline
$\gamma$ & (112) & $0.654(1)$ & ${\it 0.611(11)}$ &  \\
$\bar{\gamma}$ & (112)$[\bar{1}10]$ & ${\it 0.973(41)}$ & $0.606(5)$ &  \\
$\bar{\gamma}$ & (112)$[11\bar{1}]$ & ${\it 0.704(50)}$ & $0.550(5)$ &  \\
\end{tabular}
\end{ruledtabular}
\begin{flushleft}
{\it{$^a$ This value is for the rhcp-crystal-liquid, rather than the 
fcc-crystal-liquid interface. See \cite{davidchack_jcp125_2006} for 
details.}}
\end{flushleft}
\end{table}

In Table \ref{tab_values}, the values of $\bar{\gamma}$, as obtained
from our simulation, are given in comparison to previous simulation
results and to DFT. A direct comparison is not possible since
in DFT the {\em tensions} are calculated whereas in MC the {\em
stiffnesses} are measured. A comparison is only possible using a
tension-stiffness conversion through a least-square fit 
to the tension anisotropy expansion (\ref{equ_gamma_expansion}) and the
corresponding expression for the stiffnesses (through a combination
of (\ref{equ_gamma_expansion}) and (\ref{eqn_stiffness})), giving
the average tension $\gamma_0$ and the parameters $\epsilon_i$ ($i$=1,2,3). 
[Because the fit function (\ref{equ_gamma_expansion}) can not reproduce the inner 
anisotropy for the $(012)$ orientation (shown in the theory column of 
Tab. \ref{tab_values}) we have not taken into account the simulation data for 
the stiffnesses at the orientations $(012)$ and $(112)$ for the least-square fit. 
The different inner anisotropy therefore is not a shortcoming of DFT.]
Here, an element of uncertainty is added by the truncation
of the expansion since the single terms especially in the stiffness
expansion are not small (note also the associated error bars in converted
quantities). 

As expected for a fcc-fluid interface, DFT shows the largest
interfacial tension for the (001) interface orientation and the
lowest one for the (111) orientation, giving the tension anisotropy
$(\gamma(001)-\gamma(111))/2 = 0.025k_{\rm B}T/\sigma^2$. The average
tension $\gamma_0 = 0.664\,k_{\rm B}T/\sigma^2$ is 7.4\% higher
than that from the simulation. This deviation most likely stems from
the fact that in DFT (long-ranged) fluctuations in the interface are
averaged out. A comparison between the stiffnesses shows deviations
from up to $0.36$ for the $(112)[\bar{1}10]$ direction to less than
one percent for the $(012)[02\bar{1}]$ direction.

Previous simulations obtained the values 0.559(17)\,$k_{\rm B}T/\sigma^2$
\cite{davidchack_jcp125_2006} and 0.5610(12)\,$k_{\rm B}T/\sigma^2$
\cite{davidchack_jcp133_2010} for $\gamma_0$. These values are 10\%
smaller than our simulation results. An obvious discrepancy appears for
the (111) interface orientation where we have not observed deviations
from a fcc packing in contrast to \cite{davidchack_jcp125_2006}. Further
differences to previous simulations are the use of a different geometry
and of a rotational-invariant order parameter for the identification of
crystalline particles.

We now compare our data to real-space experiments on dispersions of
hard-sphere-like colloids. They often carry residual charges and are
polydisperse. This renders a comparison with theoretical results
on hard spheres difficult. Hitherto the interfacial tension was
indirectly measured by interpreting the probability to find small
(non-spherical) clusters in terms of classical nucleation theory
\cite{gasser_science292_2001} yielding data for a mean tension of
about $\gamma_{0}=0.1\, k_{\rm B}T/\sigma^{2}$. Clearly, given the
limitations of the hard sphere model due to particle charging
and the inherent assumption of small spherical crystalline
nuclei, this is just a rough estimate of $\gamma_{0}$.
An alternative experimental route is via the analysis of the
capillary-wave spectrum similar to what we do in our MC simulations
\cite{hernandez-guzman_pnas106_2009,ramsteiner_pre82_2010,nguyen_pre84_2011}
providing direct access to the interfacial stiffnesses. In Ref.\
\cite{hernandez-guzman_pnas106_2009}, the reported stiffness of $1.2
k_{\rm B}T/\sigma^{2}$ for an interface between a randomly stacked
hexagonal close packed (hcp) crystal and its melt is significantly higher
compared to our results which might reflect the slight charge, the limited
ensemble averaging, and an ad-hoc value for the viscosity required for
the analysis in this experiment. In Ref.~\cite{ramsteiner_pre82_2010},
the reported stiffnesses were in the range of about $(0.7-1.3) k_{\rm
B}T/\sigma^{2}$. Interestingly, the stiffness for the (011) interface
was found to be isotropic and the highest value for the stiffness
was found for the (001) orientation, oppositely to what the authors
expected and what we found for hard spheres.  This might be due to a
limited number of crystalline layers and the small gravitational length
of $\sigma/7$ which limits the thickness of the liquid. Finally,
Nguyen et al.~\cite{nguyen_pre84_2011} grew crystals of PNIPAM particles in a
temperature gradient and analyzed the capillary waves along crystal-fluid
interfaces after the removal of the temperature gradient. They measured
averaged stiffnesses for several interface orientations in the range of
$(0.19-1.13) k_{\rm B}T/\sigma^{2}$ that show the best agreement with
our results. Nevertheless, the latter experiment is also not accurate
enough to validate theory and simulation on a quantitative level and thus
more experimental explorations are required.

In conclusion, we have predicted accurate values for the anisotropic
crystal-fluid surface tensions and stiffnesses of a hard sphere
system by using both fundamental measure density functional theory
and Monte Carlo simulations. A small anisotropy in the
tensions of about 10\% was found which is, however, crucial
for the transformation to stiffnesses which differ up to a factor of
4. These predictions can help to clarify apparent discrepancies found
in real-space experiments of sterically-stabilized colloidal suspensions
\cite{gasser_science292_2001,hernandez-guzman_pnas106_2009,ramsteiner_pre82_2010,nguyen_pre84_2011}.
Since the anisotropic tensions control changes of the interfacial
shape, their precise quantitative determination help to understand
crystal nucleation \cite{auer_aps173_2005,gasser_jpcm21_2009} and
the transport of larger carriers through the interface. They may also
serve as further input to phase-field-crystal calculations which
explore solidification processes on larger length and time scales
\cite{elder_prl88_2002,emmerich_ap57_2008,tegze_prl103_2009}.

Future work should address soft interactions and attractions (as
relevant, e.g.~for colloid-polymer mixtures), in order to scan the
full range from a fluid-crystal to a vapor-crystal interface. Further
extensions can be done along similar ideas as used and proposed here for
binary mixtures. Finally the recent extension of DFT towards dynamics
for Brownian systems can be used to explore the time-dependent growth
kinetics and relaxation towards equilibrium for the hard sphere interface
\cite{teeffelen_prl100_2008}.

\begin{acknowledgments}
We thank K. Sandomirski and B. B. Laird for helpful discussions. 
This work was supported by the DFG via SPP 1296 and SFB TR6. 
\end{acknowledgments}

%\bibliography{literature}

\end{document}